\documentclass[showpacs,twocolumn,floatfix,prl]{revtex4}
\usepackage{graphicx}
\usepackage{amsfonts}
\usepackage{amsmath}
\usepackage{amssymb}
\usepackage{graphicx}%

\newcommand{\beq}{\begin{equation}}
\newcommand{\eeq}{\end{equation}}
\newcommand{\beqn}{\begin{eqnarray}}
\newcommand{\eeqn}{\end{eqnarray}}
\newcommand{\bearr}{\begin{array}}
\newcommand{\enarr}{\end{array}}

\newcommand{\eps}{\varepsilon}

\begin{document}


\title{From anomalous energy diffusion to Levy walks and heat conductivity 
\\ in one-dimensional systems}

\author{P. Cipriani$^1$, S. Denisov$^2$, and A. Politi$^{2,\dag}$}
\affiliation{$^1$ Istituto Nazionale di Ottica Applicata, \\
$^2$ Max-Planck Institut f$\ddot{u}$r Physik Komplexer Systeme,
N$\ddot{o}$thnitzer Str. 38, D-01187 Dresden, Germany}

\date{\today}
\vskip 2.cm
\begin{abstract}
The evolution of infinitesimal, localized perturbations is investigated in a
one-dimensional diatomic gas of hard-point particles (HPG) and thereby
connected to energy diffusion. As a result, a Levy walk description, which was
so far invoked to explain anomalous heat conductivity in the context of
non-interacting particles is here shown to extend to the general case of truly
many-body systems. Our approach does not only provide a firm evidence that
energy diffusion is anomalous in the HPG, but proves definitely superior to
direct methods for estimating the divergence rate of heat conductivity which
turns out to be $0.333\pm 0.004$, in perfect agreement with the dynamical
renormalization--group prediction (1/3).
\end{abstract}
\pacs{44.10.+i,05.45.Jn, 05.60.-k, 05.70.Ln}

\maketitle

After the discovery of anomalous heat conductivity in classical one-dimensional
lattice systems \cite{LLP97}, in the last years a renewed attention has been
devoted to the old problem of identifying the minimal ingredients required for
the Fourier law to be ensured. As summarized in a recent review article
\cite{LLP03}, many different models have been numerically investigated to
identify the physical conditions under which the thermal conductivity $\kappa$
diverges with the system size $L$ and, having assessed that $\kappa \approx
L^\alpha$, to determine the possibly different universality classes for the
divergence rate $\alpha$. Simultaneously, several attempts have been made to
estimate analytically the scaling behaviour of $\kappa$: self-consistent
mode-coupling theory\cite{LLP03} and the Boltzmann equation \cite{P04} suggest
that $\alpha =2/5$, while dynamical renormalization group indicates $\alpha=1/3$
\cite{NR02}. Both predictions are compatible with numerical simulations
which are, however, often affected by relatively strong finite-size corrections.
The only system where convincing results have been obtained is the Fermi, Pasta
Ulam $\beta$-model in the infinite temperature limit. Its behaviour is consistent with
$\alpha =2/5$ \cite{LLP03b}, but the simmetry of the potential casts doubts
about the generality of this model\cite{lepri}. A further simple system that
can be effectively simulated on a computer is the diatomic hard-point gas: 
there, interactions are provided by elastic collisions of point-like
particles\cite{H99}. Unfortunately, the most detailed numerical simulations
reported in the literature show a slow growth of the divergence rate with $L$,
so that some authors claim that $\alpha = 1/3$ \cite{GNY02}, while others state
that the conservative guess $\alpha =0.25$ is more realistic\cite{CP03}.
Settling this issue is not only conceptually important, but it is a necessary
requisite to later quantify finite-size corrections, a crucial issue in
applications to, e.g., carbon nanotubes, where one needs to know the prefactor
as well.

Although the problem involves intrinsically many degrees of freedom, some
researchers have tried to shed some light with reference to the simpler setup of
non-interacting particles moving along a periodic array of convex scatterers
(billiard gas channels) \cite{bill}. The absence of interactions simplifies the
task of understanding heat conductivity and allows, in particular, tracing
back heat conductivity properties to the diffusion of single particles
{\it at equilibrium}. Assuming that the mean square displacement
$\langle x^2(t)\rangle$ scales as $t^\beta$ ($\beta=1$ corresponds to normal
diffusion), it can be shown that $\alpha = \beta - 1$ \cite{DKU04}, under the
assumption that each particle exchanges energy only at the channel borders,
where thermal reservoirs operate. The limit of this approach is, on the one
hand, that the relationship between the diffusion exponent $\beta$ and the
microscopic dynamics remains to be established \cite{SZK93} and, on the other
hand, that particles do not mutually interact. Nevertheless, in this Letter we
show that upon interpreting the energy density as a pseudo-particle density,
the above reasoning can be fruitfully applied to the HPG to find a convincing
evidence that energy diffuses like in a Levy walk process\cite{ZK93}, thus
bridging two research lines that were so far basically disconnected from one
another. As a further result, we are also able to establish that in the HPG,
$\alpha = 1/3$ with a 1\% accuracy.

The model consists in a chain of $N$ point-like particles with alternating
masses $m_i$ lying on a segment of length $L$, $0\leq x_i\leq L\
(i=1,\ldots,N)$, where $m_{2i}=m$, with $(i=1,\ldots,[N/2])$, and
$m_{2i+1}=rm$, with $(i=0, \ldots,[(N-1)/2])$, where square brackets
indicate the integer part. Due to the absence of intrinsic energy- and
length-scales, we can fix them at wish. The only relevant parameter that cannot
be scaled out is the mass ratio $r$. Without loss of generality, we set the
{\sl number density} $\nu\doteq N/L = 1$, the energy per particle
$\eps \equiv\langle m_i v_i^2\rangle /2 = 1$ (which, in turn, fixes the
temperature, $k_B T\equiv 2$) and one of the two particle masses equal to 1.
The dynamics of this model is very simple, since the velocities change only
in the collisions between adjacent particles while the updating rule is
determined by the conservation laws of kinetic energy and linear momentum. By
denoting with $v_i$ ($v'_i$) the velocity before (after) a collision, the
evolution equation amounts to
\begin{equation}
v'_{i} =  v_j \pm \frac{1-r}{1+r}(v_{i}-v_j)
\label{eq:mot}
\end{equation}
where a plus (minus) sign is selected depending whether $i$ is even (odd) and
$j =i\pm1$. It is also interesting to notice that the positions $x_i$
contribute only indirectly to the evolution, by determining the collision times.
Such properties, plus the conservation of the particle {\sl ordering} along the
chain, allow simulating the dynamics with an {\sl event driven} algorithm
\cite{GNY02}, that exploits the heap structure of the future collision times.

When $r=1$, collisions just exchange particle velocities, that are, thereby,
integrals of motion. Away from $r=1$, the system is no longer integrable, but
remains non-chaotic. In fact, since the evolution equation
(\ref{eq:mot}) is linear, it describes the dynamics of an infinitesimal
perturbation $\delta v_i(t)$ as well. Therefore, the weighted Euclidean norm
\begin{equation}
 Q = \sum_i \delta_{(2)}(i,t) \equiv \sum_i m_i (\delta v_i (t))^2 
\label{eq:connorm}
\end{equation}
of a generic perturbation in tangent space is conserved for the same reason that
energy is conserved
in the phase-space dynamics, and, {\it a fortiori} the maximum Lyapunov
exponent cannot be larger than zero \cite{GNY02}.

A more detailed
characterization of the dynamics can be obtained from the spectrum $\Lambda(u)$
of convective Lyapunov exponents \cite{DK87}, which quantifies the space-time
growth rate of a perturbation $\delta_{(2)}(i,t=0) = \delta_{i,0}$ initially
localized in $i=0$ ($\delta_{i,j}$ is Kronecker $\delta$ function)
\begin{equation}
\Lambda(u)  =  \lim_{t\rightarrow \infty}
\frac{1}{2}\log \delta_{(2)}(i=ut,t) \ .
\label{eq:convec}
\end{equation}
$\Lambda(u)$ attains its maximum at $u=0$, where it coincides with the
maximum Lyapunov exponent. In a standard chaotic model, upon increasing $|u|$,
$\Lambda(u)$ decreases until it crosses 0 at $u=u_s$, a value that has been
shown to coincide with the sound velocity \cite{GHPV00}. It is then interesting
to understand how the scenario modifies in a non-chaotic model such as the HPG.
Previous studies revealed a slow convergence to the asymptotic regime both when
$r$ is close to 1 and for $r \gg 1$ \cite{GNY02}. In order to avoid such
problems we have chosen $r=3$, but we can stress that the same scenario has been
confirmed by a few simulations run for $r=2$. The data reported in
Fig.~1 shows that $\Lambda$ converges to 0 for $u<u_c\approx 1.02$
($u_c$ has been determined by fitting the long-time behaviour of the
secondary maximum of $\Lambda(u)$), while it becomes strictly negative at
larger velocities. This is analogous to harmonic chains, where the exponential
growth rate of a perturbation is strictly zero as long $|u|$ is smaller
than the sound velocity \cite{Vassalli}. We have verified that $u_c$
coincides, within the numerical error, with the sound velocity (determined from
the position of the peaks in the structure function) at the very same
temperature.

\begin{figure}[t]
\includegraphics[width=0.9\linewidth,angle=0]{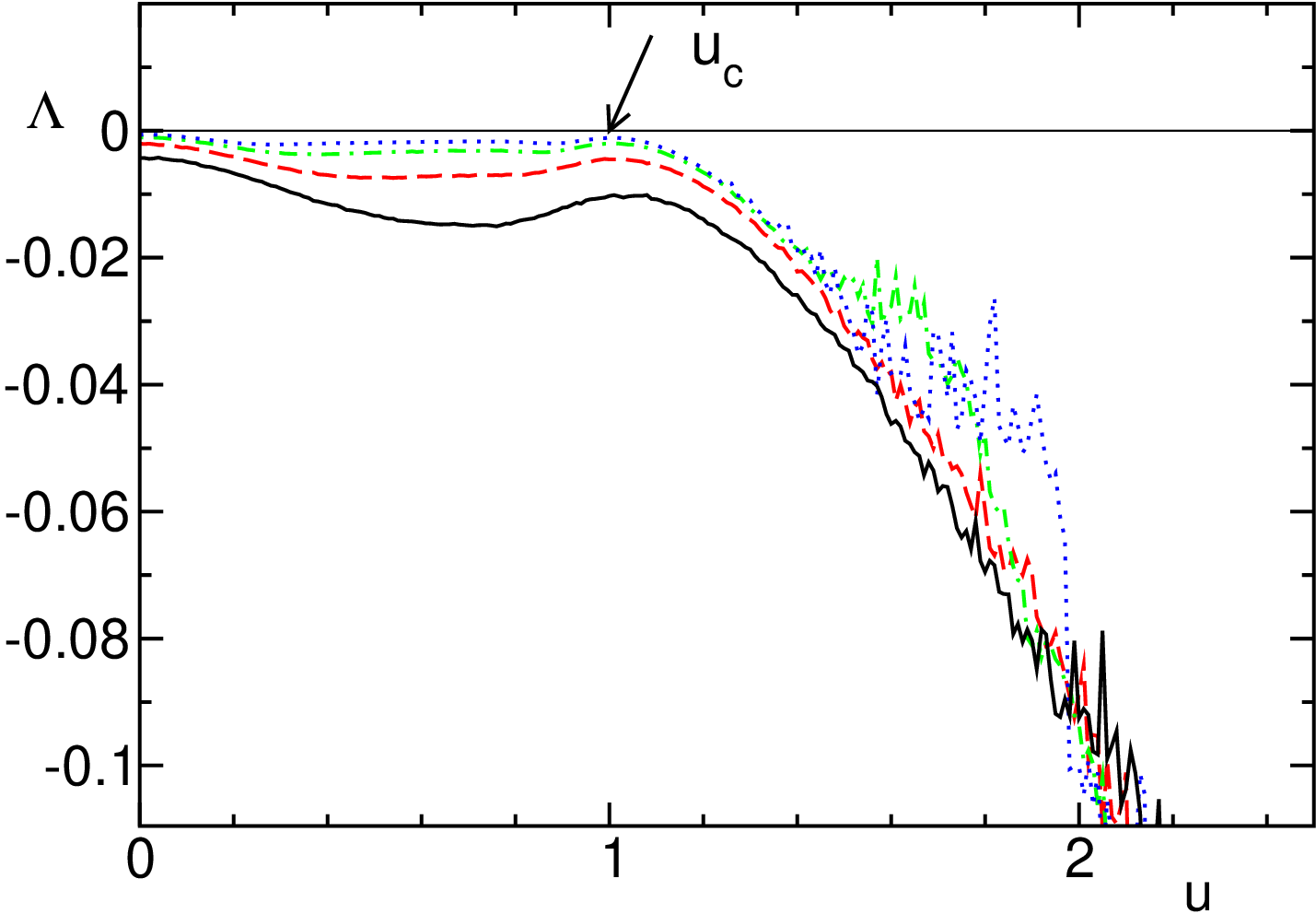}
\caption{Spectrum of comoving Lyapunov exponents at different
times for $r=3$; the chain length is N=8190. From bottom to top the
curves have been obtained by comparing perturbations ad times
(100,200), (200,400), (400,800), (800,1200) and (800,1600). The
perturbation amplitude has been averaged over $10^4$ different
realizations.} \label{fig:conv}
\end{figure}

The very observation that $\Lambda(u)= 0$ within a finite velocity range
implies a sub-exponential scaling and thereby suggests looking for a more
accurate scaling Ansatz. The power-law hypothesis
\begin{equation}
\delta_{(2)}(i,t) \approx t^{-\gamma_{1}}
\delta_{(2)}(i/t^{\gamma_{2}}), ~~~ i < u_{c}t
\label{eq:ans}
\end{equation}
proves to be the correct choice. Because of the conservation law
(\ref{eq:connorm}), we expect that temporal and spatial rates coincide with one
another, namely $\gamma_{1}=\gamma_{2}=\gamma$. The value of $\gamma$ can be
estimated from the behaviour of $\delta_{(2)}(0,t)$ which appears to follow a
clean power law (see circles in Fig.~2). A best fit (solid line)
yields $0.606\pm 0.008$ which strongly hints at $\gamma = 3/5$. By adopting
this $\gamma$ value in Eq.~(\ref{eq:ans}), the rescaled profiles $\delta_{(2)}$
are plotted in Fig.~3 at different times (without loss of generality, $Q$ is set
equal to 1). The scaling Ansatz (\ref{eq:ans}) is finally validated by the
extremely good overlap of the rescaled distributions in the interval delimited
by the secondary peaks, observed in correspondence of the sound velocity
$u_{s}$. It is interesting to notice that while the system-size $L=8190$ allows
obtaining a clean scaling behavior for the perturbation diffusion, direct 
simulations of the heat conductivity indicate that $L=30,000$ is not
large enough to obtain a looser estimate of the divergence rate
\cite{GNY02,CP03}.

\begin{figure}[t]
\includegraphics[width=0.8\linewidth,angle=0]{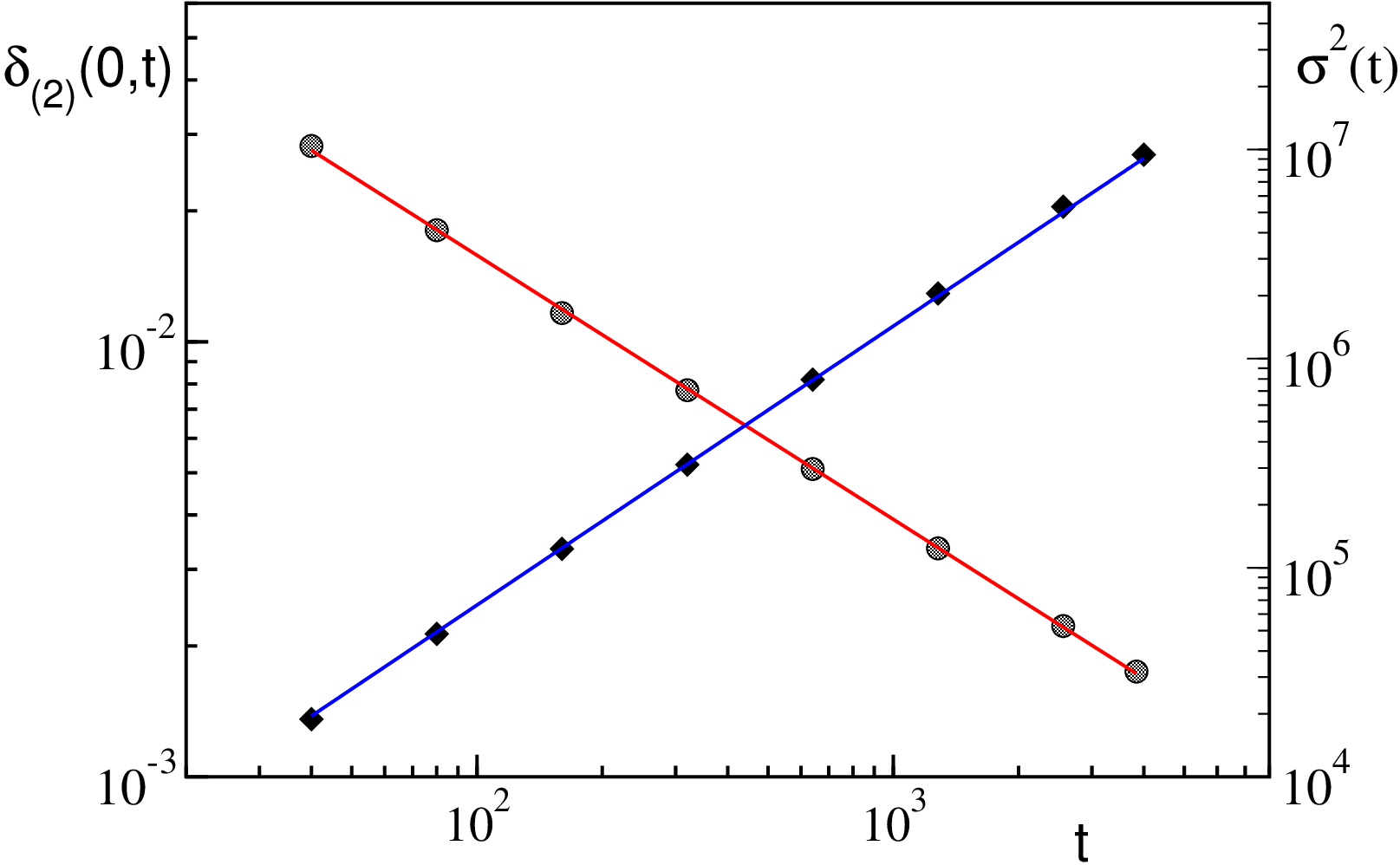}
\caption{Maximum height $\delta_{(2)}(0,t)$ of the infinitesimal perturbations
profile (circles) and mean square displacement $\sigma^{2}(t)$ (diamonds)
versus time.}
\label{alphascal8190}
\end{figure}

In view of the conservation law expressed by Eq.~(\ref{eq:connorm}), it is
tempting to interpret the perturbation profile $\delta_{(2)}(i,t)$ as a
probability density function (pdf) and thereby compare our results with those
expected for an ensemble of Levy walks \cite{ZK93}. A simple schematization
of a Levy walk consists in a particle moving ballistically in between
successive ``collisions'' whose time separation is distributed according to
a power law, $\psi(t) \propto t^{-\mu -1}$ ($\mu >0$~\cite{note1}), while
their velocity is chosen from some symmetric distribution which, in the
simplest setup, reduces to ${\cal P}(u) =(\delta(u-1) + \delta(u+1))/2$.
The propagator (the pdf to find in $x$ at time $t$, a particle initially
localized at $x=0$) of such a process is~\cite{KZ94}
\begin{equation}
  P(x,t) \propto
  \begin{cases} 
    \begin{array}{ll}
         t^{-1/\mu} \exp \left( \frac{-a x^2}{t^{2/\mu}}\right)
	                                   & |x| \lesssim t^{1/\mu} \\
	     t x^{-\mu-1}  & t^{1/\mu} \lesssim |x| <  t\\ 
        t^{1-\mu}   &|x| = t \\
         0   &|x| > t
     \end{array}
   \end{cases}
\end{equation}
It can be easily shown that, up to the length $|x|=t$, the propagator $P(x,t)$
scales as in Eq.~(\ref{eq:ans}) with the exponent $\gamma=1/\mu$. The cutoff
at $|x|=t$ is a consequence of the finite constant velocity and leads to sharp
peaks at the outermost wings (see also the inset of Fig.~3, where the propagator
for the Levy walk with $\mu=1/\gamma=5/3$ is compared with the direct simulation
of the HPG). In fact, these peaks correspond to single flights during the
observational time. The only relevant difference with the results of HPG
simulations, namely the broad secondary peaks exhibited by $\delta_{(2)}$,
disappears as soon as the $\delta$-Dirac functions in the definition of
${\cal P}(u)$ are replaced by Gaussian distributions with a suitable width
(see again the inset in Fig.~3).

As for the secondary peaks, a best fit of their height gives a decay as
$t^{-1.15\pm 0.03}$. This is to be confronted with the theoretical prediction
$t^{1 -\mu -1/\mu}$\cite{note0} which, in this case amounts to $t^{-19/15}$.
Much of the deviation is to be attributed to the not yet established asymptotic
decay in the tail.

\begin{figure}[t]
\includegraphics[width=0.98\linewidth,angle=0]{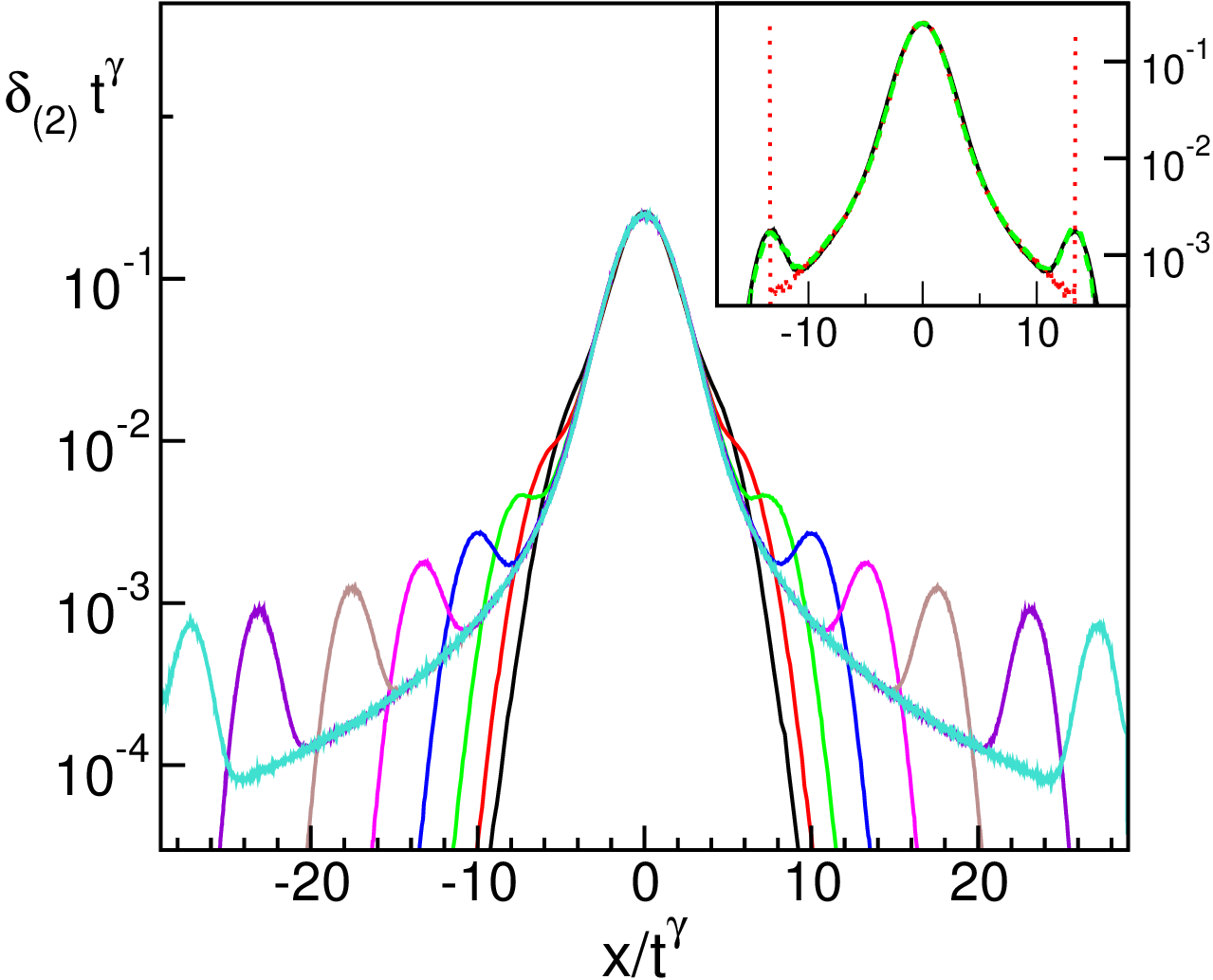}
\caption{Rescaled perturbation profiles at $t=40$, 80, 160, 320,
640, 1280, 2560, and 3840 (the width increases with time), for
$\gamma = 3/5$. The profiles have been obtained by averaging over $10^4$
realizations. In the inset, the profile at $t=640$ (solid line) is
compared with the propagators of a Levy walk for an exponent $\mu=\frac{5}{3}$
with a fixed velocity $u=1$ (dotted line) and a Gaussian distribution with
average equal to 1 and rms 0.036)(dashed line). In the last two cases, the
propagators have been obtained by averaging over $10^8$  different realizations.
\label{fig:Levy}}
\end{figure}
Let us now discuss the connection between dynamics in tangent space and energy
diffusion. The latter process can be investigated from the evolution of an
initially localized, finite perturbation $\Delta v_i(0)$,
\begin{eqnarray}
\Delta E_i(t) =  \frac{m_i}{2} (v_i(t) + \Delta v_i(t))^2 - 
  \frac{m_i}{2}v_i^2(t) =  \nonumber \\
  m_i v_i(t)\Delta v_i(t) + 
  \frac{m_i}{2}(\Delta v_i(t))^2 , \label{eq:diffu}
\end{eqnarray}
and thereby performing an ensemble-average $\langle \Delta E_i(t)\rangle$ to get
rid of statistical fluctuations. By directly computing
$\langle \Delta E_i(t) \rangle$, we found that even after averaging over
$10^{4}$ realizations, it was not possible to obtain meaningful results on time
scales shorter than those considered in the previous simulations. In fact, while
an increasing accuracy is required to resolve tinier deviations, statistical
fluctuations remain of order one at all times. Additionally, there are
fluctuation amplifications due to $\Delta v_i$ sudden jumps\cite{note3}.
However, in the limit of infinitesimal perturbations, the probability of such
jumps vanishes and the last term in the r.h.s. of Eq.~(\ref{eq:diffu}) can be
identified with $\delta_{(2)}$. As a consequence, not only the total energy of
the perturbation is connserved, but because of Eq.~(\ref{eq:connorm}), the
sums of the two terms in the r.h.s. of Eq.~(\ref{eq:diffu}) are separately
conserved. Finally, if the perturbation is randomly chosen, 
$\langle \sum m_i v_i \delta v_i \rangle$ vanishes, so that the relevant
properties of energy diffusion are captured by the pseudo-Euclidean norm of
$\delta v_i(t)$.

Therefore, from the scaling behaviour for $\delta_{(2)}$ we can conclude that 
energy diffuses anomalously in the HPG. This can be also shown more directly,
from the behaviour of the {\it mean square displacement}
$\sigma^{2}(t)= \sum_{i}i^2\delta_{(2)}(i,t)$, which is expected to scale as
$\sigma^{2}(t) \propto t^{\beta}$. A best fit of our numerical data (see
diamonds in Fig.~2) yields $\beta \approx 1.35$, a value in agreement with
Ref.~[8b]. Since in a Levy walk, the exponent $\beta$ is equal to $3-\mu$,
we expect $\beta = 4/3$, which thus confirms the validity of the Levy walk
interpretation.

Energy superdiffusion can be finally related to the anomalous behaviour of
$\kappa$, by exploiting the link established by linear response theory between
heat conductivity and the decay of spontaneous fluctuations of the energy
density \cite{Palmer}. By following Ref.~\cite{DKU04}, the divergence rate
$\alpha$ of heat conductivity can be connected to the anomalous diffusivity
exponent $\beta$ through the simple relation $\alpha=\beta-1=2-\mu=2-2/\gamma$,
which implies $\alpha =1/3$ in our case \cite{note2}, a value in perfect
agreement with the prediction of dynamical renormalization group \cite{NR02}.

The clean results summarized in the first three figures of this Letter are made
possible by the peculiarity of HPG dynamics which allows for a strong
reduction of statistical fluctuations. It should not be, however, forgotten the
absence of exponential instabilities whose presence would have obliged to
averaging over an exponentially growing number of trajectories.
Therefore, one might suspect that the Levy--walk scenario is peculiar to the
HPG, but this hypothesis contrasts the observation that the HPG scaling
behaviour coincides with the predictions based on very general hydrodynamic
arguments. On the other hand, if Levy walks generally occur in 1d systems,
evidence for this behavior should be found in, e.g., truly chaotic models:
however, the above mentioned numerical difficulties strongly suggest the need
to define a completely new protocol to answer this question.

Let us now briefly discuss the origin of the anomalous behavior.
The evolution in tangent space is fully determined by three elements: energy
plus momentum conservation and the correlations between collisions. In
Ref.~\cite{KMP82}, it was introduced a 1d model where the energy of each
particle is randomly exchanged (with one of the two neighbors) in uncorrelated
collision events. In such a simplified context it was rigorously proved that
Fourier law is satisfied. It is thus natural to ask whether the diverging heat
conductivity observed in the HPG is due by the additional conservation of
momentum (which is absent in the model of Ref.~\cite{KMP82}). By replacing the
HPG collision pattern with a set of  completely uncorrelated events, we found
a normal behaviour. Therefore, we are led to conclude that the anomaly is
entirely contained in the long-range correlations between collisions.

\acknowledgments S. Lepri is acknowledged for useful discussions. This work is
part of the project PRIN2003 {\it Order and chaos in nonlinear extended
systems} funded by MIUR Italy.


\begin{thebibliography}{1000}
\bibitem{note1} Here, we are interested in the interval $1<\mu<2$,
which corresponds to superdiffusive L$\acute{e}$vy walks \cite{ZK93}.
\bibitem{note0} By considering that the peak is positioned in
$z \approx t^{1/\mu -1}$, the proability distribution decays as
$z^{-1/\mu-1}$, the remaining area is on the order of $\zeta^{-1/\mu}$.
The result follows by reminding the scaling of the $y$ axis and assuming, as
confirmed by numerics that the peak does not broaden.
\end{thebibliography}
\end{document}